\documentclass[final,5p,times,twocolumn]{elsarticle} 
\usepackage{amssymb}

\usepackage[table,dvipsnames,x11names]{xcolor}
\usepackage[table,dvipsnames,x11names,addedmarkup=bf,addedcolor=BrickRed,deletedmarkup=sout,deletedcolor=Blue4]{changes}
\usepackage{amsmath}
\usepackage{multirow}
\usepackage{booktabs}
\usepackage{bm}
\usepackage{braket}
\usepackage{amsmath}
\usepackage{graphicx}
\usepackage[normalem]{ulem}
\biboptions{sort&compress}

\definechangesauthor[name={Xie Mengran}]{xmr}
\definechangesauthor[name={IMP}]{redauthor}
\definechangesauthor[name={Blue Author}]{blueauthor}

\usepackage[pdfstartview=FitH, colorlinks,
 linkcolor={red!60!black!},
 anchorcolor={green!80!black!},
 urlcolor={blue!80!black!},
 citecolor={blue!50!black!}]{hyperref}

\journal{Physics Letters B}

\begin{document}

\begin{frontmatter}

\title{How Threshold Effects in Spectroscopic Factors Influence Heavy-Ion Knockout Reactions}

\author[ad1,ad2]{M. R. Xie}
\author[ad1,ad2,ad3]{J.G. Li\corref{correspondence}}
\author[ad4]{C. A. Bertulani}
\author[ad1,ad2]{N. Michel}
\author[ad1,ad2]{Y. Z. Sun}
\author[ad1,ad2,ad3]{W. Zuo}

\address[ad1]{State Key Laboratory of Heavy Ion Science and Technology, Institute of Modern Physics, Chinese Academy of Sciences, Lanzhou 730000, China}

\address[ad2]{School of Nuclear Science and Technology, University of Chinese Academy of Sciences, Beijing 100049, China}

\address[ad3]{Southern Center for Nuclear-Science Theory (SCNT), Institute of Modern Physics, Chinese Academy of Sciences, Huizhou 516000, China}

\address[ad4]{Department of Physics and Astronomy, 
East Texas A\&M University, Texas 75429-3011, USA}

\cortext[correspondence]{Corresponding author. e-mail address: jianguo\_li@impcas.ac.cn (J.G. Li)}

\begin{abstract}
A two-decade‐old puzzle in heavy-ion one-nucleon knockout reactions is the strong correlation between the reduction factor $R_s=\sigma_{\rm exp}/\sigma_{\rm th}$ and the Fermi surface asymmetry $\Delta S$. Theoretical cross sections typically rely on spectroscopic factors (SFs) from shell model (SM) calculations, which neglect continuum coupling effects.
Here, we employ the Gamow shell model (GSM), which explicitly incorporates continuum coupling,  to compute SFs for $p$-shell nuclei and predict corresponding theoretical cross sections. Systematic calculations demonstrate that using GSM-derived SFs substantially reduces discrepancies between theoretical and experimental results. This improvement is particularly significant for deeply bound nucleon knockout in nuclei near the dripline, where traditional SM-based calculations fall short. As a result, using GSM SFs, the ratio $R_s$ exhibits no pronounced dependence on $\Delta S$. 
Furthermore, both the ratio of GSM SFs to SM SFs and their corresponding reaction cross sections ratios exhibit a strong $\Delta S$ dependence. 
We have also compared GSM SFs and cross sections with those from the no-core shell model calculations, giving a similar pronounced sensitivity to $\Delta S$. Detailed analysis attributes these correlations to threshold effects for SFs in weakly bound systems.
Overall, incorporating continuum coupling via GSM enhances the reliability of SF predictions for exotic, weakly bound nuclei and provides key insights toward resolving the enduring puzzle in heavy-ion knockout reactions from a nuclear structure perspective.
\end{abstract}


\end{frontmatter}

\section{Introduction}
Single-nucleon knockout reactions have emerged as a powerful tool for investigating the single-particle (s.p.) structure of unstable nuclei by extracting spectroscopic factors (SFs). These studies provide valuable insights into nuclear shell evolution, nucleon correlations, and reaction mechanisms~\cite{AUMANN2021103847, PhysRevC.93.054315}. Importantly, SFs also play a critical role in shaping our understanding and prediction of nuclear reaction rates in stellar environments~\cite{10.3389/fphy.2020.602920}.
Systematic compilations of experimental single-nucleon removal cross sections for light and medium-mass nuclei on light targets ($^{9}$Be, $^{12}$C) at intermediate energies have revealed that the reduction factor $R_s$, which is defined as the ratio of the measured cross section to its theoretical prediction, shows a strong dependence on the Fermi surface asymmetry $\Delta S$ of the projectile nucleus~\cite{PhysRevC.77.044306, PhysRevC.102.044614, PhysRevC.90.057602}.

This behavior was first identified in experiments conducted at the National Superconducting Cyclotron Laboratory (NSCL), USA, in 2004, using inverse kinematics knockout on \(^{9}\mathrm{Be}\) targets at 150~MeV/nucleon~\cite{PhysRevC.69.034311}. Subsequent experiments confirmed these trends in single-nucleon knockout reactions for light and medium-mass nuclei on \(^{9}\mathrm{Be}\) or \(^{12}\mathrm{C}\) targets, with beam energies around 100~MeV/nucleon~\cite{PhysRevC.77.044306, PhysRevC.90.057602, PhysRevLett.93.042501, PhysRevC.103.054610}. In addition, similar results were observed in experiments carried out at the Lanzhou Heavy Ion Research Facility in China, with beam energies near 250~MeV/nucleon~\cite{PhysRevC.110.014603}.   

Despite these systematic observations, the underlying cause of the \(R_s\)-\(\Delta S\) dependence remains unresolved and continues to represent a long-standing puzzle in heavy-ion-induced knockout reactions. In contrast, other experimental methods, such as transfer reactions~\cite{PhysRevLett.102.062501, PhysRevLett.111.042502, PhysRevLett.104.112701}, quasifree scattering~\cite{GOMEZRAMOS2018511, HOLL2019682,  PhysRevLett.130.172501}, 
and electron-induced \((e,e'p)\) knockout reactions~\cite{PhysRevLett.122.172502, PhysRevLett.82.4404}, have not shown any significant correlation with \(\Delta S\), highlighting the complexity and the unresolved nature of this phenomenon.

The reduction factor \(R_s\) combines inputs from both nuclear structure and reaction calculations, with knockout reaction analyses typically relying on the eikonal reaction models~\cite{BERTULANI2006372, PhysRevLett.102.232501, annurev:/content/journals/10.1146/annurev.nucl.53.041002.110406} and SFs derived from traditional shell model (SM) calculations~\cite{PhysRevC.77.044306, AUMANN2021103847, PhysRevC.90.057602, PhysRevC.103.054610}.  
Although numerous studies have focused on refining the eikonal reaction model, the persistent puzzle of the \(R_s\)-\(\Delta S\) dependence remains only partially addressed~\cite{DIAZCORTES2020135962, BERTULANI2023138250, PhysRevLett.130.172501, PhysRevLett.108.252501, GOMEZRAMOS2023138284, PhysRevC.79.064617, LI2024139143, PhysRevC.83.011601, PhysRevC.77.044306, PhysRevC.67.034317, PhysRevC.105.024613}.   
In particular, when SM SFs are used, nuclei with large $\Delta S$ values typically exhibit small $R_s$ values, often around 30\%~\cite{PhysRevC.77.044306, PhysRevC.103.054610, PhysRevC.90.057602}. These nuclei are usually located near the neutron or proton dripline, where the nuclear system is weakly bound, yet the removed nucleon is deeply bound inside the nucleus.  

In dripline nuclei, explicitly coupling to the continuum reduces the SFs of deeply bound nucleons compared to results obtained from calculations that neglect these effects~\cite{XIE2023137800, PhysRevC.104.L061301, PhysRevLett.107.032501, PhysRevC.82.044315, OKOLOWICZ2016303, PhysRevC.88.044315}. Neglecting continuum coupling therefore leads to a systematic overestimate of SFs in SM near the dripline, which in turn causes large discrepancies between theoretical and experimental one-nucleon removal cross sections and drives the observed $R_s$-$\Delta S$ dependence.

Here, we employ the Gamow shell model (GSM), which explicitly accounts for continuum coupling, to systematically calculate SFs for $p$-shell nuclei spanning a wide range of separation energies and backed by extensive experimental data. These GSM SFs are then used as nuclear structure inputs in single‐nucleon removal cross section calculations. Furthermore, we explore the correlation between \(\Delta S\) and the ratio of GSM-derived SFs to those obtained from SM calculations. This analysis provides new insight into the underlying mechanism responsible for the observed \(R_s\)-\(\Delta S\) dependence when using traditional SM SFs at the nuclear structure level.

\section{Method}
\subsection{Cross Section Calculation}
Theoretical one-nucleon removal cross section for a final state \(I^{\pi}\) at excitation energy \(E_x\) is given by:
\begin{equation}
    \sigma_{\text{th}}(I^{\pi}) = f_{\text{CoM}} \, C^2S(I^{\pi}, j) \, \sigma_{\text{sp}}(j^{\pi}, S_n + E_x),
    \label{EQ}
\end{equation}
where \(f_{\text{CoM}}\) is the center-of-mass motion correction, taken as \((A/(A-1))^N\) in traditional SM calculations~\cite{annurev:/content/journals/10.1146/annurev.nucl.53.041002.110406, PhysRevC.10.543}, and equal to 1 in GSM and no-core shell model (NCSM) calculations. The term \(C^2S\) represents the SF, which includes isospin coupling~\cite{PhysRevC.10.543}, while \(\sigma_{\text{sp}}\) denotes the s.p. cross sections, calculated using the CNOK code based on the Glauber model~\cite{SUN2023108726}.  

In the calculation of \(\sigma_{\text{sp}}\), nuclear densities (target and residual nuclei) and root-mean-square (rms) radii of the removed single nucleon used for the reaction calculations are obtained from Hartree-Fock-Bogoliubov (HFB) calculations employing four Skyrme interactions: SLY4, SLY5, SIII, and SKM*~\cite{BENNACEUR200596}. The final cross sections are averaged over the results from these interactions. Detailed results of the s.p. cross sections and other relevant information are provided in the Supplementary Material.

For a single final state \(I^{\pi}\), corresponding to the ground state of the residual nucleus, the asymmetry parameter \(\Delta S\) is defined as
$\Delta S  = S_n - S_p$ for neutron removal, and $\Delta S = S_p - S_n$ for proton removal, where $S_p$ and $S_n$ are the proton and neutron separation energies, respectively~\cite{PhysRevC.90.057602, PhysRevC.77.044306}.
When multiple final states of the residue are populated, \(\Delta S\) is replaced by a weighted average of the effective separation energies, where the weights are given by the calculated partial cross sections \(\sigma_{\text{th}}(I^{\pi})\)~\cite{PhysRevC.90.057602}, based on the GSM results presented in this work. In such cases, the total theoretical cross section is obtained by summing the partial cross sections of all populated final states:
$
\sigma_{\text{th, total}} = \sum_{I^{\pi}} \sigma_{\text{th}}(I^{\pi}).
$

\subsection{Spectroscopic Factors and Overlap Function}
The theoretical SF is defined as the norm of the overlap function:
\begin{equation}
    C^2S_{\ell j} = \int_0^{+\infty} O_{\ell j}^2(r) \, dr, 
    \label{SF}
\end{equation}
where the overlap functions are given by:
\begin{equation}
    O_{\ell j}(r) = \frac{1}{\sqrt{2J_A + 1}} \sum_n \langle \Psi_A^{J_A} \Vert a_{n\ell j}^+ \Vert \Psi_{A-1}^{J_{A-1}}\rangle u_n^{(\ell j)}(r),
    \label{overlap_function}
\end{equation}
with $\ell$ and $j$ representing orbital and total angular momentum of the partial wave, respectively. 
Here, $J_A$ and $J_{A-1}$ are the total angular momenta of the nuclear wave functions $\mid \Psi_A^{J_A} \rangle$ and $\mid \Psi_{A-1}^{J_{A-1}} \rangle$, corresponding to the $A$ and $A-1$ nucleon systems. The functions $u_n^{(\ell j)}(r)$ are the radial components of the one-body harmonic oscillator (HO) basis in SM and NCSM, while the one-body Berggren basis in GSM.

\subsection{Gamow Shell Model}
The GSM extends the traditional SM into the complex momentum ($k$) plane~\cite{Michel:2002, Michel:2021}, utilizing the Berggren basis~\cite{Berggren:1968, Michel:2002, Michel:2021}.
For the Berggren basis, each partial wave $\ell,j$ spans as:
\begin{equation}
\sum_n u_{n}^{(\ell j)}(r) u_{n}^{(\ell j)}(r') + \int_{L^+} u_{k}^{(\ell j)}(r)  u_{k}^{(\ell j)}(r') ~dk = \delta(r - r'), 
\label{Berggren}
\end{equation}
where $n$ enumerates the bound and resonance states of the considered partial wave, while $L^+$
is the complex contour of scattering states, which encompasses the resonance states present in the discrete sum. 
This framework makes GSM particularly powerful for studying weakly bound nuclei and their continuum coupling.
In practical calculations, continuous states on the contour $L^+$ are discretized by the Gauss-Legendre quadrature~\cite{Michel_2009} with sufficient points in each partial wave to ensure the reliability of the results.

In the GSM framework, SFs are independent of the choice of the s.p. Berggren basis~\cite{Michel:2021}. Although SFs calculated within the GSM framework are inherently complex quantities~\cite{ PhysRevC.75.031301, Xie2023, XIE2023137800}, this study focuses solely on their real parts.
The nucleus is described as a system of valence nucleons interacting outside an inert closed core ($^4$He in this work), following the Cluster Orbital Shell Model (COSM) framework~\cite{PhysRevC.38.410}.


In the GSM calculations, the core-valence interaction is modeled using a finite-depth Woods–Saxon potential, while the residual two-body interaction is described by the Furutani-Horiuchi-Tamagaki (FHT) effective nucleon-nucleon force~\cite{10.1143/PTP.62.981}. The parameters of these interactions and the details of the model space used for the \(p\)-shell nuclei calculations are provided in Refs.~\cite{PhysRevC.96.054316, XIE2023137800, Xie2023}.

To enhance computational efficiency and improve the convergence of SF calculations, particularly for nuclei with mass numbers \(A = 9\) and \(A = 10\), natural orbitals are employed in our model space. These orbitals are obtained as eigenstates of the scalar density matrix of the many-body GSM wave function constructed in the Berggren basis, thereby effectively capturing a significant portion of the correlation and interaction strength in these systems. 
As a concrete example, consider the GSM calculations for $^9$Be. First, calculations are carried out using the Berggren basis, allowing at most two particles in the continuum, to generate the natural orbital basis. The dimension is on the order of $10^8$. Subsequently, GSM calculations are performed within the constructed natural orbital basis, now allowing up to three particles in the continuum, with the dimension again being on the order of $10^8$. For comparison, the corresponding traditional SM calculation for $^9$Be within the $p$-shell valence space involves only about $10^2$ basis states.

\subsection{Shell Model}
SM calculations are typically performed using a HO basis, with an inert core—usually a doubly magic nucleus—assumed to remain inactive. The many-body correlations among nucleons are incorporated through configuration mixing within a truncated model space,  which restricts active nucleons to occupy one or two major shells, known as the valence space. The effective interaction in the valence space is generally phenomenological, constructed by fitting experimental data such as energy levels and transition rates.

In the present work, SM calculations are carried out using the KSHELL code~\cite{SHIMIZU2019372}. We choose $^{4}$He as the inert core and take the $0p_{3/2}$ and $0p_{1/2}$ orbitals as the valence space for both protons and neutrons. The effective interaction employed is the phenomenological Cohen-Kurath interaction~\cite{COHEN19671}, which has been widely used and validated in $p$-shell nuclei.

\subsection{No-Core Shell Model}
The \textit{ab initio} NCSM has achieved significant success in describing light nuclei and ranks among the most microscopic many-body approaches currently available for studying light nuclear systems. Unlike traditional SM, NCSM starts from realistic nuclear interactions and does not assume an inert core.
The NCSM calculation is formulated in an HO basis and truncated by a maximum total excitation quantum number $N_{\rm max}$~\cite{BARRETT2013131} to control the rapid growth in model space dimension. In this work, we use the realistic Daejeon16 nuclear interaction~\cite{SHIROKOV201687}, which is derived from the chiral two-nucleon interaction at next-to-next-to-next-to-leading order through the similarity renormalization group evolution and unitary transformation techniques. 
It provides an excellent description of the properties of light nuclei without the need to include three-body forces. 
Due to computational constraints, we apply a truncation of $N_{\rm max} = 10$ for nuclei with $A \le 6$, and $N_{\rm max} = 8$ for heavier $p$-shell nuclei. At the same time, we adopt an oscillator frequency of $\hbar w = 15$ MeV.

\section{Results}

\begin{figure}[!htb]
    \centering
    \includegraphics[width=0.95\columnwidth]{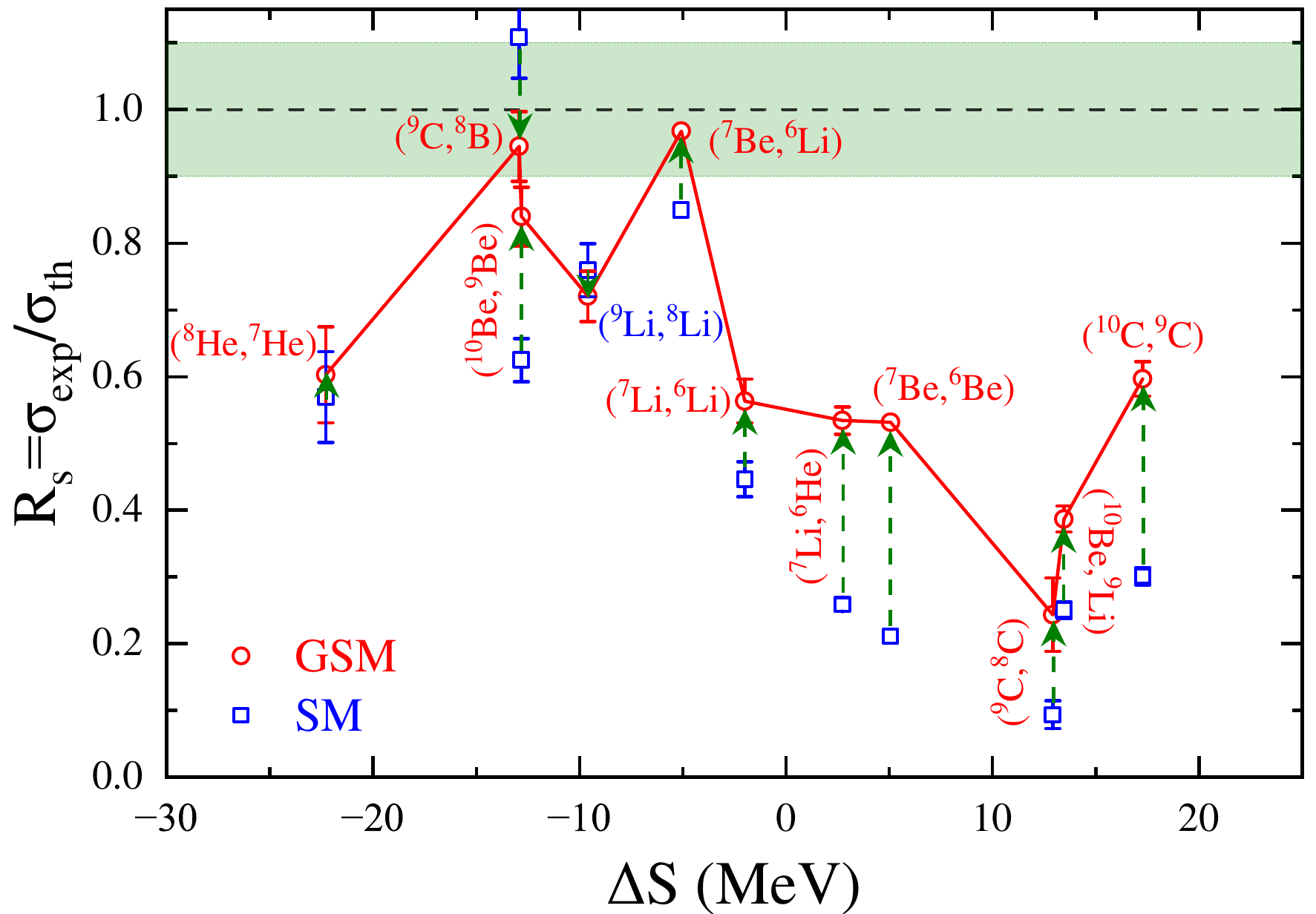}
    \caption{The reduction factor \(R_s\) as a function of the asymmetry parameter \(\Delta S\). Experimental inclusive nucleon removal cross sections, \(\sigma_{\text{exp}}\), are taken from Refs.~\cite{PhysRevLett.106.162502, PhysRevLett.102.232501, PhysRevC.86.024315, PhysRevC.102.044614, MARKENROTH2001462, AKSYUTINA2009191, 10.1063/1.4909557}. The theoretical cross sections, \(\sigma_{\text{th}}\), are computed by combining reaction model calculations with nuclear structure inputs derived from both GSM and SM SFs.} 
    \label{sigma}
\end{figure}

SFs obtained from GSM calculations are used to compute theoretical single-nucleon removal cross sections \(\sigma_{\text{th}}\) for those \(p\)-shell nuclei with available experimental data.
For comparison, cross section calculations are also performed using SFs derived from traditional SM calculations. The reduction factor, defined as \(R_s = \sigma_{\text{exp}} / \sigma_{\text{th}}\), quantifies the discrepancy between theoretical calculations and measured inclusive cross sections, as shown in Fig.~\ref{sigma}. The uncertainties presented include both experimental uncertainties and the theoretical spread in \(\sigma_{\text{th}}\), the latter arising from the use of HFB calculations with different Skyrme interactions. Among these two sources, the experimental uncertainties dominate.

For knockout reaction calculations using SFs derived from the SM, a significant overestimation of theoretical cross sections compared to experimental data is observed, resulting in systematically small values of \(R_s\). 
This overestimation is particularly pronounced for reactions with large \(\Delta S\) values, underscoring the strong correlation between \(R_s\) and \(\Delta S\). Such cases typically involve projectile nuclei that are either bound or weakly bound (e.g., \({}^{7,10}\mathrm{Be}\), \({}^{10}\mathrm{C}\)), and residual nuclei that are either weakly bound (e.g., \({}^{6}\mathrm{He}\), \({}^{9}\mathrm{Li}\), \({}^{9}\mathrm{C}\)) or unbound states (e.g., \({}^{6}\mathrm{Be}\), \({}^{8}\mathrm{C}\)) located near or beyond the dripline.
For proton removal from \({}^{7}\mathrm{Li}\) and \({}^{10}\mathrm{Be}\), the calculated cross sections exceed experimental measurements by factors of approximately 3.86 and 3.99, respectively. 
Similarly, neutron knockout cross sections from \({}^{9,10}\mathrm{C}\) and \({}^{7}\mathrm{Be}\) are also significantly overestimated.

\begin{figure}[!htb]
    \centering
    \includegraphics[width=0.95\columnwidth]{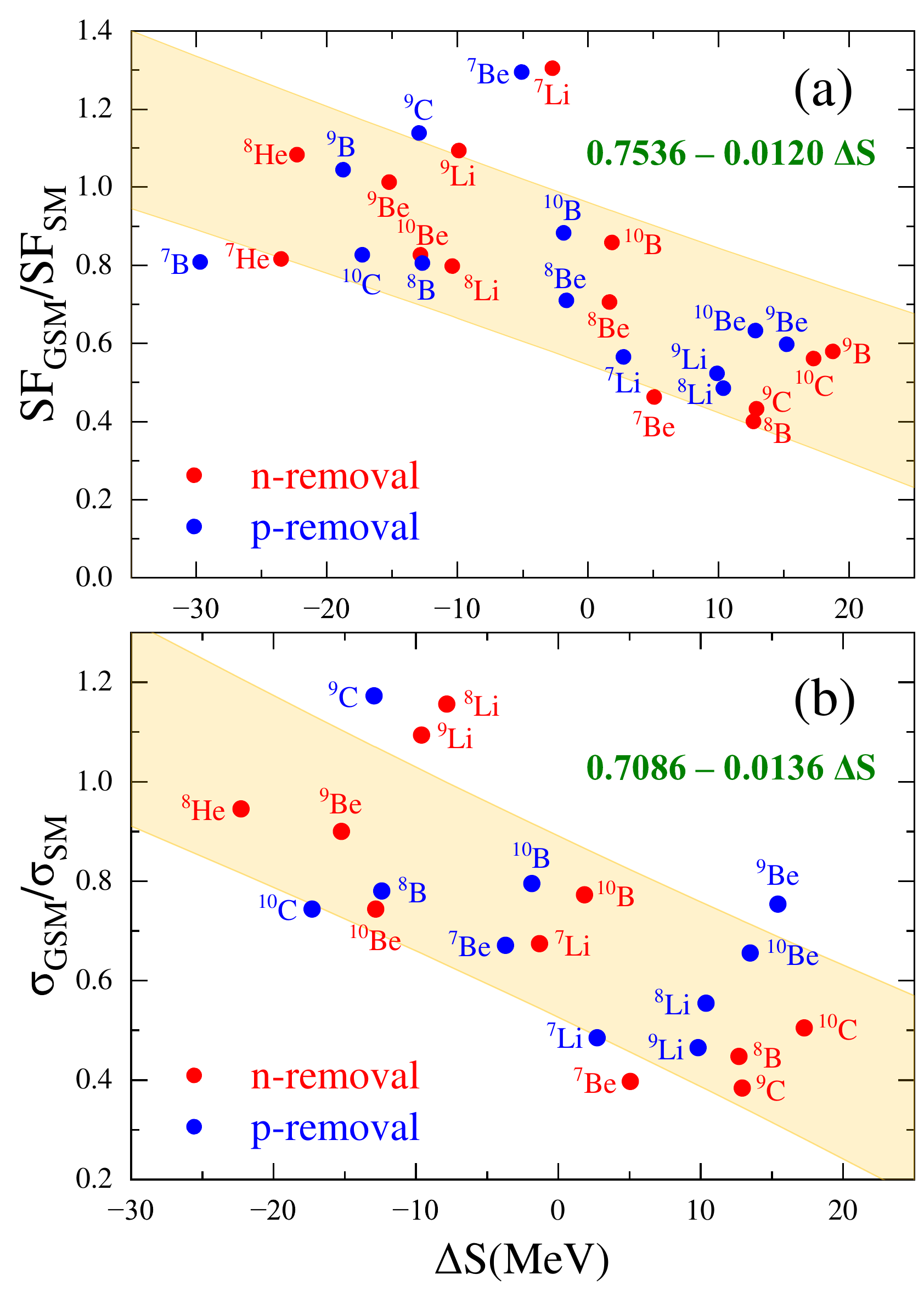}
    \caption{Panels (a) and (b) show the ratios \(\mathrm{SF}_{\mathrm{GSM}}/\mathrm{SF}_{\mathrm{SM}}\) and \(\sigma_{\mathrm{GSM}}/\sigma_{\mathrm{SM}}\) plotted as functions of the asymmetry parameter \(\Delta S\), respectively. The shaded region indicates the \(1\sigma\) uncertainty from the linear fit.} 
    \label{RS_th}
\end{figure}

In contrast, the use of GSM-derived SFs significantly improves the agreement between calculated and experimental cross sections, particularly for nuclei with large \(\Delta S\). For instance, GSM-based calculations for proton knockout from \({}^{7}\mathrm{Li}\) and neutron knockout from \({}^{7}\mathrm{Be}\) and \({}^{10}\mathrm{C}\) show \(R_s\) values increasing from around 0.3 (in SM calculations) to approximately 0.6, demonstrating a clear improvement. Furthermore, GSM-based calculations substantially reduce and flatten the \(R_s\)-\(\Delta S\) dependence, in stark contrast to the pronounced trend observed in SM-based results. Although the GSM-based calculations show significant improvements over the traditional SM, a sizable dispersion in $R_s\approx 0.3–0.7$ persists in the region of large $\Delta S$.

These discrepancies are driven by differences in the SFs obtained from the two models. For the \(({}^{10}\mathrm{C},{}^{9}\mathrm{C})\) reaction, GSM predicts an SF of 0.973, compared to the SM value of 1.735, resulting in a much more accurate prediction of the experimental cross section.
Similarly, for the \(({}^{9}\mathrm{C}, {}^{8}\mathrm{C})\) reaction at 66.8 MeV/nucleon, GSM yields an SF of 0.374, which is much closer to the experimental cross section~\cite{PhysRevC.102.044614}, in contrast to the SM calculation of 0.886.

Within the GSM framework, the weakly bound or unbound nature of residual nuclei is explicitly treated, resulting in consistently smaller SFs for deeply bound nucleons. In contrast, SM calculations, which rely on an HO basis with an infinitely deep potential well, are unable to adequately describe the properties of these nuclei near the dripline~\cite{XIE2023137800, Xie2023}.

The enhanced accuracy of knockout cross section predictions using GSM SFs, compared to those obtained from SM calculations and experimental measurements, highlights the strong sensitivity of knockout reactions to underlying nuclear structure effects. This motivates a direct comparison between SFs from GSM and SM, focusing in particular on the \(p_{3/2}\) partial wave, which dominates the ground state of the studied \(p\)-shell isotopes.

Fig.~\ref{RS_th} (a) presents the ratio \(\mathrm{SF}_{\mathrm{GSM}}/\mathrm{SF}_{\mathrm{SM}}\) as a function of \(\Delta S\), with the shaded band representing the \(1\sigma\) uncertainty of linear regression. The observed trend follows the relation:
$\mathrm{SF}_{\mathrm{GSM}}/\mathrm{SF}_{\mathrm{SM}} = 0.7536 - 0.0120 \Delta S$. 
This behavior closely resembles the experimentally observed trend for the ratio \(\mathrm{SF}_{\mathrm{exp}}/\mathrm{SF}_{\mathrm{SM}}\) in neutron knockout reactions within the argon isotope chain, given by:
$\mathrm{SF}_{\mathrm{exp}}/\mathrm{SF}_{\mathrm{SM}} = 0.55 - 0.0175 \Delta S$~\cite{PhysRevLett.131.212503}.
Similar results are also obtained in the shell model embedded in the continuum calculations~\cite{OKOLOWICZ2016303} and in SFs derived from overlap functions calculated via solving the inhomogeneous equation~\cite{ PhysRevC.88.044315}. 
By embedding the continuum coupling at the single-particle level, the GSM propagates it consistently to the many-body space and thereby captures more long-range correlations beyond the reach of bound-state bases in the traditional SM. The obtained $\mathrm{SF}_{\mathrm{GSM}}/\mathrm{SF}_{\mathrm{SM}}$ suggests that continuum-induced long-range correlations may play a significant role in shaping the observed $R_s$–$\Delta S$ systematics.
Moreover, at \(\Delta S = 0\),  \(\mathrm{SF}_{\mathrm{GSM}}/\mathrm{SF}_{\mathrm{SM}}\) $\approx$ 0.730, demonstrating significant improvement in describing SF quenching. This result explains why cross section calculations based on GSM SFs are in much closer agreement with experimental data than those using SM SFs and why GSM-based calculations substantially reduce the \(R_s\)-\(\Delta S\) dependence.

\begin{figure}[!htb]
    \centering
    \includegraphics[width=0.95\linewidth]{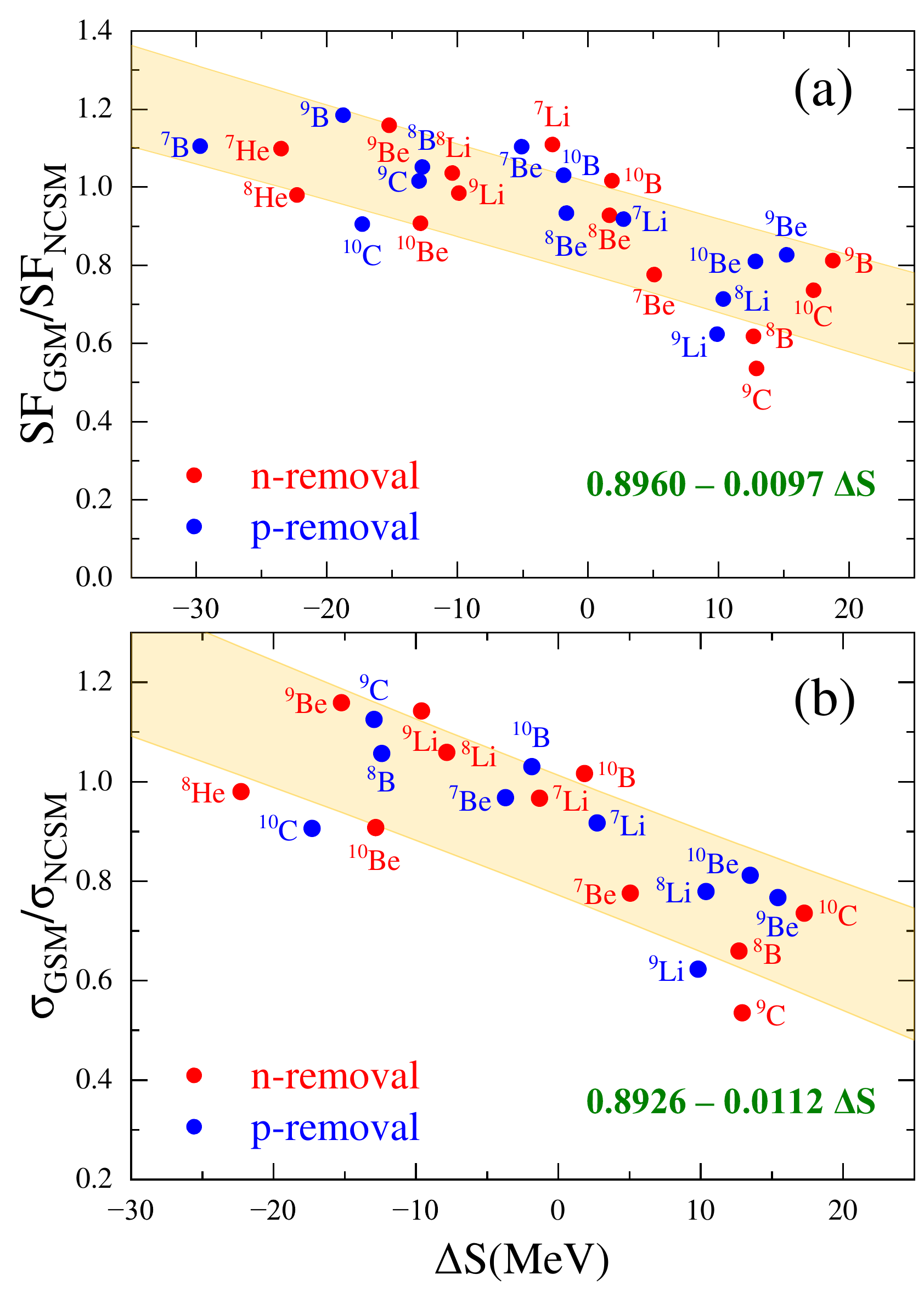}
    \caption{Similar to Fig.~\ref{RS_th}, but with $\rm SF_{NCSM}$ and $\rm \sigma_{NCSM}$ rather than $\rm SF_{SM}$ and $\rm \sigma_{SM}$.}
    \label{NCSM-Re}
\end{figure}

Based on the calculated SFs, we have calculated the inclusive cross sections for \(p\)-shell knockout reactions, including all particle-bound final states associated with the \(p_{3/2}\) and \(p_{1/2}\) partial waves, using both SM and GSM SFs. For consistency, the analysis is focused on reactions at an incident energy of 100~MeV/nucleon with a \({}^{9}\mathrm{Be}\) target. However, it is worth noting that these conclusions remain robust under varying experimental conditions, as demonstrated in single-proton removal from the carbon isotope chain at 240~MeV/nucleon on a carbon target~\cite{PhysRevC.110.014603}.  

The results for the ratio \(\sigma_{\mathrm{GSM}}/\sigma_{\mathrm{SM}}\) are shown in Fig.~\ref{RS_th} (b). A clear dependence of \(\sigma_{\mathrm{GSM}}/\sigma_{\mathrm{SM}}\) on \(\Delta S\) is observed. A linear fit yields the relation:
$\sigma_{\mathrm{GSM}}/\sigma_{\mathrm{SM}} = 0.7086 - 0.0136 \Delta S.$
This trend closely mirrors the systematic analysis of intermediate-energy single-nucleon removal cross sections, where the experimental ratio is given by:
$\sigma_{\mathrm{exp}}/\sigma_{\mathrm{SM}} = 0.61 - 0.016 \Delta S$~\cite{PhysRevC.103.054610}.
Notably, the slope of \(\sigma_{\mathrm{GSM}}/\sigma_{\mathrm{SM}}\) (\(-0.0136\)) is slightly smaller than that of \(\sigma_{\mathrm{exp}}/\sigma_{\mathrm{SM}}\) (\(-0.0160\)) reported in Ref.~\cite{PhysRevC.103.054610}, indicating a reduced dependence on \(\Delta S\) when GSM SFs are employed in knockout reaction calculations, as illustrated in Fig.~\ref{sigma}.  

Furthermore, comparing the intercepts of \(\sigma_{\mathrm{exp}}/\sigma_{\mathrm{SM}}\) and \(\sigma_{\mathrm{GSM}}/\sigma_{\mathrm{SM}}\) yields a ratio of approximately 0.861. This suggests that replacing SM SFs with GSM SFs can significantly improve the description of cross section quenching. A similar improvement may also be achievable for cross section quenching observed in transfer reactions~\cite{PhysRevLett.111.042502} by employing GSM SFs.

We also performed the SFs calculations using the state-of-the-art \textit{ab initio} NCSM, and carried out parallel cross section calculations within the same reaction framework. Compared to the SM, NCSM incorporates a broader configuration space~\cite{PhysRevC.110.054301}, but without including the continuum effects.
Fig.~\ref{NCSM-Re} (a) and (b) display the ratios of GSM to NCSM SFs and cross sections plotted as functions of the asymmetry parameter $\Delta S$, respectively.
While NCSM-based calculations significantly improve the absolute cross sections, both the calculated \(\mathrm{SF}_{\mathrm{GSM}}/\mathrm{SF}_{\mathrm{NCSM}}\) and \(\sigma_{\mathrm{GSM}}/\sigma_{\mathrm{NCSM}}\) still exhibit a pronounced sensitivity to the Fermi surface asymmetry $\Delta S$.
The results further strengthen the reliability of the above conclusion, in which continuum coupling plays a crucial role in the SF calculations at the nuclear structure level.
Furthermore, consistent with the theoretical SFs for $^{15}$C reported in Ref.~\cite{JIANG2025139789}, the results show that at large negative $\Delta S$ the GSM SFs are systematically larger than their NCSM counterparts. A natural interpretation is that GSM underrepresents short-range correlations, leading to larger SFs, whereas NCSM, with extensive multi-configuration mixing in a large no-core space, partially incorporates these correlations responsible for quenching.

\begin{figure}[h]
    \centering
    \includegraphics[width=0.95\columnwidth]{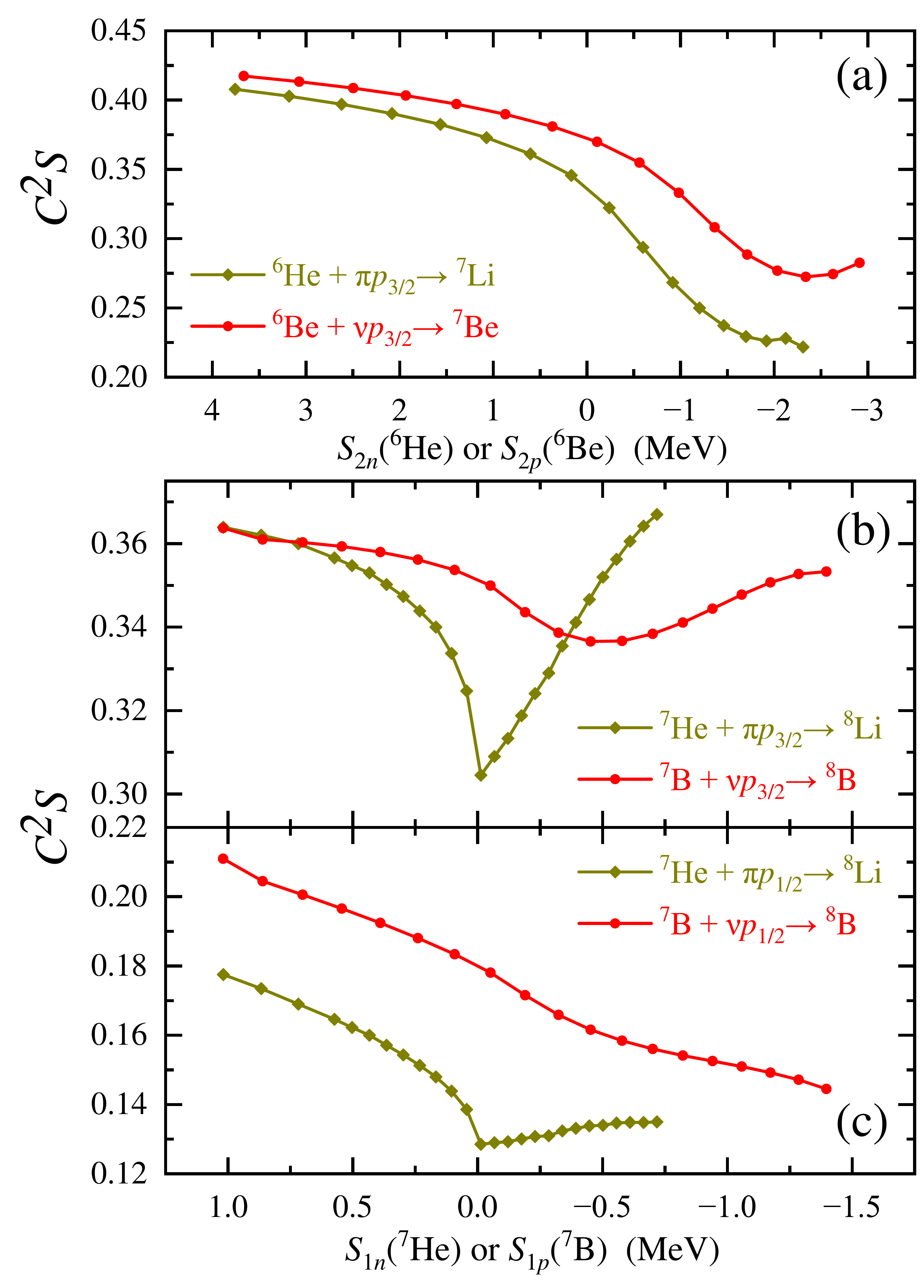}
    \caption{SFs as functions of separation energy.  
    (a) Proton SF of the ground state of \({}^{7}\mathrm{Li}\) and neutron SF of its mirror nucleus \({}^{7}\mathrm{Be}\), plotted against the two-nucleon separation energy \(S_{2n/2p}\). 
    (b) Proton SF of the ground state of \({}^{8}\mathrm{Li}\) and neutron SF of its mirror nucleus \({}^{8}\mathrm{B}\) with $p_{3/2}$ partial waves and (c) corresponding results with $p_{1/2}$ partial waves, shown as functions of the single-nucleon separation energy \(S_{1n/1p}\).}
    \label{V0}
\end{figure}

To explore the mechanisms responsible for the reduction of SFs associated with deeply bound nucleons in projectile isotopes near the dripline, we performed GSM calculations for \(A=7\) (\({}^{7}\mathrm{Li}\) and \({}^{7}\mathrm{Be}\)) and \(A=8\) (\({}^{8}\mathrm{Li}\) and \({}^{8}\mathrm{B}\)) isotopes. In these calculations, the Hamiltonian was adjusted by varying the depth of the Woods–Saxon potential within the GSM framework, allowing us to tune the nucleon separation energies of the residual isotopes from deeply bound to unbound states. This approach enabled us to investigate the impact of threshold effects in the residual nuclei (\({}^{6,7}\mathrm{He}\), \({}^{6}\mathrm{Be}\) and \({}^{7}\mathrm{B}\)) on the SFs of deeply bound nucleons in the projectile nuclei, thereby enhancing our understanding of their structural evolution near the dripline.

Fig.~\ref{V0} (a) presents GSM results for SFs of deeply bound protons in the \(p_{3/2}\) partial wave for the ground state of \({}^{7}\mathrm{Li}\), plotted as a function of the two-neutron separation energy \(S_{2n}\) of \({}^{6}\mathrm{He}\). The corresponding mirror case, involving neutron SFs in \({}^{7}\mathrm{Be}\) versus the two-proton separation energy \(S_{2p}\) of $^6$Be, is also shown. As \(S_{2n/2p}\) decreases, the residual nucleus transitions from a bound to a weakly bound state, resulting in a pronounced reduction of the SF. In particular, the SF values decrease sharply as the system approaches the particle emission threshold. Beyond this threshold, the SFs reach a minimum and display an inflection point. Interestingly, the inflection point occurs at a larger value of \(S_{2p}\) in \({}^{7}\mathrm{Be}\) neutron SF compared to \(S_{2n}\) in \({}^{7}\mathrm{Li}\) proton SF, a shift attributed to the higher Coulomb barrier present in proton-rich nuclei relative to their mirror neutron-rich counterparts.

Fig.~\ref{V0} (b) displays proton SF of the \({}^{8}\mathrm{Li}\) ground state and neutron SFs of its mirror nucleus \({}^{8}\mathrm{B}\) for the $p_{3/2}$ partial waves, while Fig.~\ref{V0} (c) shows the corresponding $p_{1/2}$ partial wave results, both plotted as functions of the single-nucleon separation energies ($S_{1n}$).
In the case of \({}^{8}\mathrm{Li}\), SF shows a pronounced threshold effect as the GSM Hamiltonian is adjusted to approach \(S_{1n} = 0\). 

In contrast, for proton-rich nucleus \({}^{8}\mathrm{B}\), the threshold effect is less pronounced. This is consistent with the fact that the Wigner-cusp phenomenon is strongest for low-angular-momentum waves in neutron systems without a Coulomb barrier, while the presence of a Coulomb barrier in proton-rich systems suppresses this effect. Similarly, the inflection point observed near the single-neutron threshold $S_{1n}$ of $^6$He in \({}^{7}\mathrm{Li}\) proton SF  calculation also follows the Wigner threshold behavior.

These results demonstrate the ability of many-body open quantum system calculations to accurately capture Wigner-cusp behavior and threshold effects arising from continuum coupling, thus validating the GSM framework. As discussed earlier, the improved agreement of deeply bound nucleon knockout cross section calculations near the dripline can be attributed to the lower SF values obtained from GSM compared to those from traditional SM. In the GSM approach, SFs exhibit strong, nonlinear variations near particle emission thresholds, emphasizing the critical importance of including threshold effects when extracting SFs in dripline nuclei. In contrast, traditional SM calculations lack these dynamic features, as their SF values remain fixed and do not account for the dramatic structural variations observed near thresholds.

\section{Summary}

This study addresses the long-standing puzzle of heavy-ion-induced knockout reactions by employing SFs derived from the GSM as nuclear structure inputs for cross section calculations. 
Our results demonstrate that SM-based calculations systematically overestimate experimental cross sections for nuclei with large \(\Delta S\), leading to a strong and persistent correlation between the reduction factor $R_s$ and $\Delta S$. In contrast, the GSM, which incorporates continuum coupling through the Berggren basis, significantly improves agreement with experimental data, particularly for weakly bound nuclei near the dripline, thereby greatly reducing the $R_s$-$\Delta S$ dependence.
Moreover, the ratio $\rm SF_{GSM}/SF_{SM}$ and 
$\rm SF_{GSM}/SF_{NCSM}$ revealing a strong dependence on \(\Delta S\), which further underscores the role of continuum effects.  
Detailed GSM analyses attribute this behavior to threshold effects in dripline nuclei, which are not captured by traditional SM and NCSM. These findings suggest that GSM-based calculations provide a potential resolution to the observed \(R_s\)-\(\Delta S\) dependence from a nuclear structure perspective.

Nevertheless, a residual weak \(R_s\)-\(\Delta S\) dependence remains, indicating that complementary refinements to reaction models—beyond structural corrections—are necessary to resolve this discrepancy fully.
Moreover, SM SFs are used as nuclear structure inputs in transfer reactions~\cite{PhysRevLett.110.122503, PhysRevLett.131.212503} and \((p,pN)\) reaction~\cite{HOLL2019682, GOMEZRAMOS2018511} calculations similarly exhibit weak \(R_s\)-\(\Delta S\) correlation. Replacing SM SFs with GSM SFs in these reaction models could further reduce, or potentially eliminate, the \(R_s\)-\(\Delta S\) dependence.

\textit{Acknowledgments.}~
This work has been supported by the National Key R\&D Program of China under Grant Nos. 2023YFA1606403, 2024YFE0109800, and 2024YFE0109802; the National Natural Science Foundation of China under Grant Nos.  12205340, 12175281, 12347106, and 12121005;  the Gansu Natural Science Foundation under Grant No. 25JRRA467;  the Strategic Priority Research Program of Chinese Academy of Sciences under Grant No. XDB34000000;  C.A.B acknowledges support by the U.S. DOE Grant No. DE-FG02-08ER41533. The numerical calculations in this paper have been done on Hefei advanced computing center.

\section*{References}

\bibliographystyle{elsarticle-num_noURL}
\bibliography{Ref}

\end{document}



\title{Supplemental Material for: [How Threshold Effects in Spectroscopic Factors Influence Heavy-Ion Knockout Reactions]}

\author[ad1,ad2]{M. R. Xie}
\author[ad1,ad2,ad3]{J.G. Li\corref{correspondence}}
\author[ad4]{C. A. Bertulani}
\author[ad1,ad2]{N. Michel}
\author[ad1,ad2]{Y. Z. Sun}
\author[ad1,ad2,ad3]{W. Zuo}

\address[ad1]{State Key Laboratory of Heavy Ion Science and Technology, Institute of Modern Physics, Chinese Academy of Sciences, Lanzhou 730000, China}

\address[ad2]{School of Nuclear Science and Technology, University of Chinese Academy of Sciences, Beijing 100049, China}

\address[ad3]{Southern Center for Nuclear-Science Theory (SCNT), Institute of Modern Physics, Chinese Academy of Sciences, Huizhou 516000, China}

\address[ad4]{Department of Physics and Astronomy, 
East Texas A\&M University, Texas 75429-3011, USA}

\cortext[correspondence]{Corresponding author. e-mail address: jianguo\_li@impcas.ac.cn (J.G. Li)}


\maketitle

\section{HFB Charge Radius Calculations vs. Experimental Data}
The calculated and experimental charge radii of $p$-shell nuclei are listed in Table~\ref{rms}, using four different Skyrme interactions.

\begin{table*}[h]
\centering
\caption{The Hartree-Fock-Bogoliubov method is utilized to compute the nuclear charge radii (in fm) of $p$-shell nuclei using four Skyrme parameterizations, with corresponding experimental measurements provided for comparison. The experimental data, denoted as "Expt.", were sourced from the National Nuclear Data Center (NNDC).}
\label{rms}
    \begin{tabular}{cccccc}
    \hline\hline
    Nucleus & SLY4 & SLY5 & SIII & SKM* & Expt. \\ \hline
    $^6$He  & 2.136 & 2.137 & 2.066 & 2.134 & 2.066 \\
    $^7$He  & 2.143 & 2.145 & 2.079 & 2.139 &  \\
    $^8$He  & 2.145 & 2.149 & 2.091 & 2.143 & 1.924 \\
    $^6$Li  & 2.335 & 2.336 & 2.254 & 2.321 & 2.589 \\
    $^7$Li  & 2.325 & 2.324 & 2.251 & 2.310 & 2.444 \\
    $^8$Li  & 2.321 & 2.321 & 2.256 & 2.306 & 2.339 \\
    $^9$Li  & 2.318 & 2.317 & 2.264 & 2.306 & 2.245 \\
    $^6$Be  & 2.581 & 2.590 & 2.476 & 2.531 &  \\
    $^7$Be  & 2.492 & 2.495 & 2.404 & 2.464 & 2.646 \\
    $^8$Be  & 2.459 & 2.459 & 2.383 & 2.434 &  \\
    $^9$Be  & 2.443 & 2.442 & 2.378 & 2.419 & 2.519 \\
 $^{10}$Be  & 2.433 & 2.432 & 2.606 & 2.411 & 2.355 \\
    $^7$B   & 2.763 & 2.779 & 2.653 & 2.691 &  \\
    $^8$B   & 2.616 & 2.623 & 2.527 & 2.582 &  \\
    $^9$B   & 2.561 & 2.564 & 2.487 & 2.533 &  \\
 $^{10}$B   & 2.534 & 2.535 & 2.473 & 2.508 & 2.428 \\
    $^8$C   & 2.918 & 2.947 & 2.817 & 2.835 &  \\
    $^9$C   & 2.713 & 2.728 & 2.633 & 2.685 &  \\
 $^{10}$C   & 2.640 & 2.648 & 2.575 & 2.618 &  \\
    \hline\hline
    \end{tabular}
\end{table*}

\section{Systematic Nucleon Knockout in the p-Shell: Cross Sections, Spectroscopic Factors, and HFB Inputs}

The data required for the calculation of one-nucleon knockout reaction cross sections in the $p$-shell have been systematically compiled. This compilation includes the spectroscopic factor results calculated using different theoretical models, the single-particle cross sections provided by the reaction models, and the nuclear densities and root-mean-square (rms) values (in fm) used in these calculations.

\begin{table*}[h]
\centering
\caption{The results of the neutron knockout reaction in the $p$-shell have been systematically compiled. The analysis assumes that the incident nucleus collides with a $^9$Be target at an energy of 100 MeV/nucleon. Theoretical cross sections $\sigma_{th}$ are derived using the spectroscopic factors (denoted as $C^2S$) from each model, incorporating contributions from both the $p_{3/2}$ and $p_{1/2}$ partial waves. $\sigma_{sp}$ is the single-particle cross section.
Here, $S_n^*$ represents the effective single-neutron separation energy, while $\Delta S$ denotes the corresponding Fermi surface asymmetry.
The rms values refer to the single-nucleon removal radii. The nuclear densities and rms values (in fm) utilized in these calculations are obtained from the Hartree-Fock-Bogoliubov method employing the SLY4 interaction.
}
\label{SLY41}
\resizebox{\textwidth}{!}{
\begin{tabular}{ccccccccccccccc}
\hline\hline
& & &  & \multicolumn{5}{c}{$p_{3/2}$} & & \multicolumn{5}{c}{$p_{1/2}$} \\ \cline{5-9} \cline{11-15}
$A-1$ & $A$ & $S_{n}*$ & $\Delta S$ & {$rms$} & {$\sigma_{sp}$} & $C^2S_{\rm GSM}$ & $C^2S_{\rm SM}$ & $C^2S_{\rm NCSM}$ & & {$rms$} & {$\sigma_{sp}$} & $C^2S_{\rm GSM}$ & $C^2S_{\rm SM}$ & $C^2S_{\rm NCSM}$ \\ \hline
$^7$He$(3/2^-)$ & $^8$He$(0^+)$ & $2.535$ & $-22.281$ & 2.961 & 63.709 & $3.251$ & $3.009$ & 3.317 & & * & * & * & * & *\\
$^6$Li$(1^+)$ & $^7$Li$(3/2^-)$ & $7.251$ & $-2.723$ & 2.648 & 64.154 & $0.522$ & $0.400$ & 0.471 &  & 2.682 & 65.361 & $0.187$ & 0.228 & 0.193 \\
$^6$Li$(3^+)$ &  & 9.437 & $-0.537$ &  & 64.370 & 0.425 & 0.683 & 0.439 & & * & * & * & * & * \\
$^6$Li$(0^+)$ &  & 10.814 & 0.840 &  & 64.292 & 0.180 & 0.333 & 0.200 & & * & * & * & * & * \\
$^6$Li$(2^+)$ &  & 11.563 & 1.589 &  & 64.280 & $0.034$ & 0.000 & 0.058 & & 2.682 & 65.495 & $0.045$ & 0.125 & 0.079 \\
{$^7$Li$(3/2^-)$} & {$^8$Li$(2^+)$} & $2.033$ & $-10.383$ & 2.726 & 57.588 & $0.804$ & $1.008$ & 0.776 &  & 2.766 & 58.734 & $0.102$ & 0.077 & 0.066 \\ 
$^7$Li$(1/2^-)$ &  & 2.511 & $-9.905$ &  & 58.360 & 0.284 & 0.155 & 0.232 & & * & * & * & * & * \\
$^7$Li$(7/2^-)$ &  & 6.663 & $-5.753$ &  & 58.539 & 0.116 & 0.226 & 0.189 & & * & * & * & * & *  \\
$^7$Li$(5/2^-)$ &  & 8.713 & $-3.703$ &  & 58.589 & 0.552 & 0.014 & 0.462 & & 2.766 & 59.963 & 0.087 & 0.001 & 0.112  \\
$^8$Li$(2^+)$ & $^9$Li$(3/2^-)$ & $4.062$ & $-9.882$ & 2.780 & 53.854 & $0.883$ & $0.807$  & 0.897 &  & 2.993 & 60.439 & $0.121$ & 0.007 & 0.041 \\
$^8$Li$(1^+)$ &  & 5.043 & $-8.901$ & & 53.518 & $0.403$ & 0.133 & 0.307 & & 2.993 & 60.578 & $0.007$ & 0.190 & 0.002  \\
$^6$Be$(0^+)$ & $^7$Be$(3/2^-)$ & $10.677$ & $5.070$ & 2.521 & 58.625 & $0.308$ & $0.665$ & 0.397 &  & * & * & * & * & *\\
{$^8$Be$(0^+)$} & {$^9$Be$(3/2^-)$} & $1.665$ & $-15.221$ & 2.661 & 49.430 & $0.575$ & $0.568$ & 0.496 &  & * & * & * & * & * \\
$^9$Be$(3/2^-)$ & $^{10}$Be$(0^+)$ & $6.812$ & $-12.824$ & 2.705 & 46.643 & $1.948$ & $2.357$ & 2.147 & & * & * & * & * & * \\
$^7$B$(3/2^-)$ & $^8$B$(2^+)$ & $12.826$ & $12.690$ & 2.514 & 49.695 & $0.290$ & $0.725$ & 0.470 & & 2.529 & 50.161 & 0.103 & 0.044 & 0.127 \\
$^9$B$(3/2^-)$ & $^{10}$B$(3^+)$ & $8.437$ & $1.850$ & 2.629 & 44.395 & $0.515$ & $0.600$ & 0.507 &  & * & * & * & * & * \\
$^8$C$(0^+)$ & $^9$C$(3/2^-)$ & $14.225$ & $12.925$ & 2.510 & 42.783 & $0.374$ & $0.866$ & 0.699 &  & * & * & * & * & * \\
$^9$C$(3/2^-)$ & $^{10}$C$(0^+)$ & $21.284$ & $17.277$ & 2.566 & 41.319 & $0.973$ & $1.735$ & 1.323 &  & * & * & * & * & *\\ 
\hline\hline
\end{tabular}}
\end{table*}

\begin{table*}[]
\centering
\caption{Similar to Table~\ref{SLY41}, but for proton knockout reaction.}
\resizebox{\textwidth}{!}{
\begin{tabular}{ccccccccccccccc}
\hline\hline
& & &  & \multicolumn{5}{c}{$p_{3/2}$} & & \multicolumn{5}{c}{$p_{1/2}$} \\ \cline{5-9} \cline{11-15}
$A-1$ & $A$ & $S_{p}*$ & $\Delta S$ & {$rms$} & {$\sigma_{sp}$} & $C^2S_{\rm GSM}$ & $C^2S_{\rm SM}$ & $C^2S_{\rm NCSM}$ & & {$rms$} & {$\sigma_{sp}$} & $C^2S_{\rm GSM}$ & $C^2S_{\rm SM}$ & $C^2S_{\rm NCSM}$ \\ \hline
$^6$He$(0^+)$ & $^7$Li$(3/2^-)$ & $9.974$ & $2.723$ & 2.530 & 63.944 & $0.376$ & $0.665$ & 0.410 &  & * & * & * & * & * \\
$^7$He$(3/2^-)$ & $^8$Li$(2^+)$ & $12.416$ & $10.383$ & 2.519 & 54.989 & $0.352$ & $0.725$ & 0.492 &  & 2.534 & 55.471 & 0.134 & 0.044 & 0.132 \\
$^8$He$(0^+)$ & $^9$Li$(3/2^-)$ & $13.944$ & $9.882$ & 2.523 & 48.462 & $0.453$ & $0.866$ & 0.727 &  & * & * & * & * & * \\
$^6$Li$(1^+)$ & $^7$Be$(3/2^-)$ & $5.607$ & $-5.070$ & 2.680 & 68.650 & $0.518$ & $0.400$ & 0.469 &  & 2.716 & 69.956 & 0.189 & 0.228 & 0.196 \\
$^6$Li$(3^+)$ &  & 7.793 & $-2.884$ &  & 68.828 & 0.426 & 0.683 & 0.430 & & * & * & * & * & * \\
$^6$Li$(0^+)$ &  & 9.170 & $-1.507$ &  & 68.890 & 0.180 & 0.333 & 0.199 & & * & * & * & *  & * \\
$^6$Li$(2^+)$ &  & 9.919 & $-0.758$ &  & 68.950 & 0.039 & 0.000 & 0.057 & & 2.716 & 70.260 & 0.031 & 0.125 & 0.080 \\
$^8$Li$(2^+)$ & $^9$Be$(3/2^-)$ & $16.886$ & $15.221$ & 2.591 & 51.373 & $0.568$ & $0.951$ & 0.687 &  & 2.605 & 51.823 & 0.139 & 0.075 & 0.156 \\
$^8$Li$(1^+)$ &  & 17.617 & 15.952 &  & 51.368 & 0.106 & 0.055 & 0.226 & & 2.605 & 51.818 & 0.175 & 0.086 & 0.220 \\
$^9$Li$(3/2^-)$ & $^{10}$Be$(0^+)$ & $19.636$ & $12.824$ & 2.574 & 45.618 & $1.098$ & $1.735$ & 1.356 &  & * & * & * & * & * \\
$^9$Li$(1/2^-)$ &  & 22.327 & 15.515 & * & * & * & * & * & & 2.585 & 45.927 & 0.332 & 0.255 & 0.421     \\
{$^7$Be$(3/2^-)$} & {$^8$B$(2^+)$} & $0.136$ & $-12.690$ & 2.775 & 69.242 & $0.812$ & $1.008$ & 0.772 &  & 2.820 & 71.436 & 0.115 & 0.077 & 0.074 \\
$^7$Be$(1/2^-)$ & & 0.565 & $-12.261$ & & 59.472 & 0.278 & 0.155 & 0.230 &  & * & * & * & * & * \\
$^7$Be$(7/2^-)$ & & 4.706 & $-8.120$ & & 62.607 & 0.108 & 0.226 & 0.171 &  & * & * & * & * & * \\
$^9$Be$(3/2^-)$ & $^{10}$B$(3^+)$ & $6.587$ & $-1.850$ & 2.650 & 47.967 & $0.530$ & $0.600$ & 0.514 &  & * & * & * & * & *  \\
{$^8$B$(2^+)$} & {$^9$C$(3/2^-)$} & $1.300$ & $-12.925$ & 2.845 & 54.223 & $0.919$ & $0.807$ & 0.905 &  & 3.100 & 62.370 & 0.137 & 0.007 & 0.045 \\
$^9$B$(3/2^-)$ & $^{10}$C$(0^+)$ & $4.007$ & $-17.277$ & 2.747 & 50.391 & $1.948$ & $2.357$ & 2.150 &  & * & * & * & * & * \\
\hline\hline
\end{tabular}}
\end{table*}

\begin{table*}[]
\centering
\caption{The results of the neutron knockout reaction in the $p$-shell have been systematically compiled. The analysis assumes that the incident nucleus collides with a $^9$Be target at an energy of 100 MeV/nucleon. Theoretical cross sections $\sigma_{th}$ are derived using the spectroscopic factors (denoted as $C^2S$) from each model, incorporating contributions from both the $p_{3/2}$ and $p_{1/2}$ partial waves. $\sigma_{sp}$ is the single-particle cross section.
Here, $S_n^*$ represents the effective single-neutron separation energy, while $\Delta S$ denotes the corresponding Fermi surface asymmetry.
The rms values refer to the single-nucleon removal radii. The nuclear densities and rms values (in fm) utilized in these calculations are obtained from the Hartree-Fock-Bogoliubov method employing the SLY5 interaction.}
\label{SLY51}
\resizebox{\textwidth}{!}{
\begin{tabular}{ccccccccccccccc}
\hline\hline
& & &  & \multicolumn{5}{c}{$p_{3/2}$} & & \multicolumn{5}{c}{$p_{1/2}$} \\ \cline{5-9} \cline{11-15}
$A-1$ & $A$ & $S_{n}*$ & $\Delta S$ & {$rms$} & {$\sigma_{sp}$} & $C^2S_{\rm GSM}$ & $C^2S_{\rm SM}$ & $C^2S_{\rm NCSM}$ & & {$rms$} & {$\sigma_{sp}$} & $C^2S_{\rm GSM}$ & $C^2S_{\rm SM}$ & $C^2S_{\rm NCSM}$ \\ \hline
$^7$He$(3/2^-)$ & $^8$He$(0^+)$ & $2.535$ & $-22.281$ & 2.999 & 64.795 & $3.251$ & $3.009$ & 3.317 &  & * & * & * & * & * \\
$^6$Li$(1^+)$ & $^7$Li$(3/2^-)$ & $7.251$ & $-2.723$ & 2.652 & 64.351 & $0.522$ & $0.400$ & 0.471 &  & 2.674 & 65.134 & $0.187$ & 0.228 & 0.193 \\
$^6$Li$(3^+)$ &  & 9.437 & $-0.537$ &  & 64.569 & 0.425 & 0.683 & 0.439 & & * & * & * & * & * \\
$^6$Li$(0^+)$ &  & 10.814 & 0.840 &  & 64.492 & 0.180 & 0.333 & 0.200 & & * & * & * & * & * \\
$^6$Li$(2^+)$ &  & 11.563 & 1.589 &  & 64.480 & $0.034$ & 0.000 & 0.058 & & 2.674 & 65.265 & $0.045$ & 0.125 & 0.079 \\
{$^7$Li$(3/2^-)$} & {$^8$Li$(2^+)$} & $2.033$ & $-10.383$ & 2.735 & 57.874 & $0.804$ & $1.008$ & 0.776 &  & 2.758 & 58.511 & $0.102$ & 0.077 & 0.066 \\ 
$^7$Li$(1/2^-)$ &  & 2.511 & $-9.905$ &  & 58.653 & 0.284 & 0.155 & 0.232 & & * & * & * & * & * \\
$^7$Li$(7/2^-)$ &  & 6.663 & $-5.753$ &  & 58.883 & 0.116 & 0.226 & 0.189 & & * & * & * & * & *  \\
$^7$Li$(5/2^-)$ &  & 8.713 & $-3.703$ &  & 58.933 & 0.552 & 0.014 & 0.462 & & 2.758 & 59.725 & 0.087 & 0.001 & 0.112  \\
$^8$Li$(2^+)$ & $^9$Li$(3/2^-)$ & $4.062$ & $-9.882$ & 2.797 & 53.950 & $0.883$ & $0.807$ & 0.897 &  & 2.959 & 59.252 & $0.121$ & 0.007 & 0.041 \\
$^8$Li$(1^+)$ &  & 5.043 & $-8.901$ & & 54.038 & $0.403$ & 0.133 & 0.307 & & 2.959 & 59.399 & $0.007$ & 0.190  & 0.002 \\
$^6$Be$(0^+)$ & $^7$Be$(3/2^-)$ & $10.677$ & $5.070$ & 2.518 & 58.441 & $0.308$ & $0.665$ & 0.397 &  & * & * & * & * & * \\
{$^8$Be$(0^+)$} & {$^9$Be$(3/2^-)$} & $1.665$ & $-15.221$ & 2.664 & 49.562 & $0.575$ & $0.568$ & 0.496 &  & * & * & * & * & * \\
$^9$Be$(3/2^-)$ & $^{10}$Be$(0^+)$ & $6.812$ & $-12.824$ & 2.716 & 47.004 & $1.948$ & $2.357$ & 2.147 &  & * & * & * & * & *  \\
$^7$B$(3/2^-)$ & $^8$B$(2^+)$ & $12.826$ & $12.690$ & 2.511 & 49.455 & $0.290$ & $0.725$ & 0.470 &  & 2.527 & 49.931 & 0.103 & 0.044 & 0.127  \\
$^9$B$(3/2^-)$ & $^{10}$B$(3^+)$ & $8.437$ & $1.850$ & 2.629 & 44.430 & $0.515$ & $0.600$ & 0.507 &  & * & * & * & * & * \\
$^8$C$(0^+)$ & $^9$C$(3/2^-)$ & $14.225$ & $12.925$ & 2.505 & 42.358 & $0.374$ & $0.866$ & 0.699 &  & * & * & * & * & * \\
$^9$C$(3/2^-)$ & $^{10}$C$(0^+)$ & $21.284$ & $17.277$ & 2.562 & 41.117 & $0.973$ & $1.735$ & 1.323 &  & * & * & * & * & * \\ 
\hline\hline
\end{tabular}}
\end{table*}

\begin{table*}[]
\centering
\caption{Similar to Table~\ref{SLY51}, but for proton knockout reaction.}
\resizebox{\textwidth}{!}{
\begin{tabular}{ccccccccccccccc}
\hline\hline
& & &  & \multicolumn{5}{c}{$p_{3/2}$} & & \multicolumn{5}{c}{$p_{1/2}$} \\ \cline{5-9} \cline{11-15}
$A-1$ & $A$ & $S_{p}*$ & $\Delta S$ & {$rms$} & {$\sigma_{sp}$} & $C^2S_{\rm GSM}$ & $C^2S_{\rm SM}$ & $C^2S_{\rm NCSM}$ & & {$rms$} & {$\sigma_{sp}$} & $C^2S_{\rm GSM}$ & $C^2S_{\rm SM}$ & $C^2S_{\rm NCSM}$ \\ \hline
$^6$He$(0^+)$ & $^7$Li$(3/2^-)$ & $9.974$ & $2.723$ & 2.528 & 63.835 & $0.376$ & $0.665$ & 0.410 &  & * & * & * & * & * \\
$^7$He$(3/2^-)$ & $^8$Li$(2^+)$ & $12.416$ & $10.383$ & 2.515 & 54.693 & $0.352$ & $0.725$ & 0.492 &  & 2.532 & 55.238 & 0.134 & 0.044 & 0.132 \\
$^8$He$(0^+)$ & $^9$Li$(3/2^-)$ & $13.944$ & $9.882$ & 2.523 & 47.613 & $0.453$ & $0.866$ & 0.727 &  & * & * & * & * & * \\
$^6$Li$(1^+)$ & $^7$Be$(3/2^-)$ & $5.607$ & $-5.070$ & 2.684 & 68.850 & $0.518$ & $0.400$ & 0.469 &  & 2.708 & 69.720 & 0.189 & 0.228 & 0.196 \\
$^6$Li$(3^+)$ &  & 7.793 & $-2.884$ &  & 69.031 & 0.426 & 0.683 & 0.430 & & * & * & * & *  & * \\
$^6$Li$(0^+)$ &  & 9.170 & $-1.507$ &  & 69.094 & 0.180 & 0.333 & 0.199 & & * & * & * & * & * \\
$^6$Li$(2^+)$ &  & 9.919 & $-0.758$ &  & 69.112 & 0.039 & 0.000 & 0.057 & & 2.708 & 70.020 & 0.031 & 0.125 & 0.080 \\
$^8$Li$(2^+)$ & $^9$Be$(3/2^-)$ & $16.886$ & $15.221$ & 2.589 & 51.291 & $0.568$ & $0.951$ & 0.687 &  & 2.600 & 51.644 & 0.139 & 0.075 & 0.156 \\
$^8$Li$(1^+)$ &  & 17.617 & $15.952$ &  & 51.285 & 0.106 & 0.055 & 0.226 & & 2.600 & 51.639 & 0.175 & 0.086 & 0.220 \\
$^9$Li$(3/2^-)$ & $^{10}$Be$(0^+)$ & $19.636$ & $12.824$ & 2.572 & 45.465 & $1.098$ & $1.735$ & 1.356 &  & * & * & * & * & * \\
$^9$Li$(1/2^-)$ &  & 22.327 & 15.515 & * & * & * & * & * & & 2.583 & 45.773 & 0.332 & 0.255 & 0.421      \\
{$^7$Be$(3/2^-)$} & {$^8$B$(2^+)$} & $0.136$ & $-12.690$ & 2.784 & 69.276 & $0.812$ & $1.008$ & 0.772 &  & 2.811 & 71.481 & 0.115 & 0.077 & 0.074 \\
$^7$Be$(1/2^-)$ & & 0.565 & $-12.261$ & & 59.788 & 0.278 & 0.155 & 0.230 &  & * & * & * & * & * \\
$^7$Be$(7/2^-)$ & & 4.706 & $-8.120$ & & 62.957 & 0.108 & 0.226 & 0.171 &  & * & * & * & * & * \\
$^9$Be$(3/2^-)$ & $^{10}$B$(3^+)$ & $6.587$ & $-1.850$ & 2.650 & 47.998 & $0.530$ & $0.600$ & 0.514 &  & * & * & * & * & * \\
{$^8$B$(2^+)$} & {$^9$C$(3/2^-)$} & $1.300$ & $-12.925$ & 2.866 & 56.920 & $0.919$ & $0.807$ & 0.905 &  & 3.060 & 63.264 & 0.137 & 0.007 & 0.045 \\
$^9$B$(3/2^-)$ & $^{10}$C$(0^+)$ & $4.007$ & $-17.277$ & 2.756 & 50.309 & $1.948$ & $2.357$ & 2.150 &  & * & * & * & * & * \\
\hline\hline
\end{tabular}}
\end{table*}

\begin{table*}[]
\centering
\caption{The results of the neutron knockout reaction in the $p$-shell have been systematically compiled. The analysis assumes that the incident nucleus collides with a $^9$Be target at an energy of 100 MeV/nucleon. Theoretical cross sections $\sigma_{th}$ are derived using the spectroscopic factors (denoted as $C^2S$) from each model, incorporating contributions from both the $p_{3/2}$ and $p_{1/2}$ partial waves. $\sigma_{sp}$ is the single-particle cross section. Here, $S_n^*$ represents the effective single-neutron separation energy, while $\Delta S$ denotes the corresponding Fermi surface asymmetry.
The rms values refer to the single-nucleon removal radii. The nuclear densities and rms values (in fm) utilized in these calculations are obtained from the Hartree-Fock-Bogoliubov method employing the SIII interaction.}
\label{SIII1}
\resizebox{\textwidth}{!}{
\begin{tabular}{ccccccccccccccc}
\hline\hline
& & &  & \multicolumn{5}{c}{$p_{3/2}$} & & \multicolumn{5}{c}{$p_{1/2}$} \\ \cline{5-9} \cline{11-15}
$A-1$ & $A$ & $S_{n}*$ & $\Delta S$ & {$rms$} & {$\sigma_{sp}$} & $C^2S_{\rm GSM}$ & $C^2S_{\rm SM}$ & $C^2S_{\rm NCSM}$ & & {$rms$} & {$\sigma_{sp}$} & $C^2S_{\rm GSM}$ & $C^2S_{\rm SM}$ & $C^2S_{\rm NCSM}$ \\ \hline
$^7$He$(3/2^-)$ & $^8$He$(0^+)$ & $2.535$ & $-22.281$ & 2.848
 & 64.903 & $3.251$ & $3.009$ & 3.317 &  & * & * & * & * & *\\
$^6$Li$(1^+)$ & $^7$Li$(3/2^-)$ & $7.251$ & $-2.723$ & 2.534
 & 65.045 & $0.522$ & $0.400$ & 0.471 &  & 2.547
 & 66.296 & $0.187$ & 0.228 & 0.193 \\
$^6$Li$(3^+)$ &  & 9.437 & $-0.537$ &  & 65.291 & 0.425 & 0.683 & 0.439 & & * & * & * & * & * \\
$^6$Li$(0^+)$ &  & 10.814 & $0.840$ &  & 65.219 & 0.180 & 0.333 & 0.200 & & * & * & * & * & * \\
$^6$Li$(2^+)$ &  & 11.563 & $1.589$ &  & 65.211 & $0.034$ & 0.000 & 0.058 & & 2.547 & 66.470 & $0.045$ & 0.125 & 0.079 \\
{$^7$Li$(3/2^-)$} & {$^8$Li$(2^+)$} & $2.033$ & $-10.383$ & 2.617 & 58.389 & $0.804$ & $1.008$ & 0.776 &  & 2.630 & 59.574 & $0.102$ & 0.077 & 0.066 \\ 
$^7$Li$(1/2^-)$ &  & 2.511 & $-9.905$ &  & 59.225 & 0.284 & 0.155 & 0.232 & & * & * & * & * & * \\
$^7$Li$(7/2^-)$ &  & 6.663 & $-5.753$ &  & 59.540 & 0.116 & 0.226 & 0.189 & & * & * & * & * & *  \\
$^7$Li$(5/2^-)$ &  & 8.713 & $-3.703$ &  & 59.615 & 0.552 & 0.014 & 0.462 & & 2.630 & 61.404 & 0.087 & 0.001 & 0.112  \\
$^8$Li$(2^+)$ & $^9$Li$(3/2^-)$ & $4.062$ & $-9.882$ & 2.686
 & 54.805 & $0.883$ & $0.807$ & 0.897 &  & 2.912
 & 61.620 & $0.121$ & 0.007 & 0.041 \\
$^8$Li$(1^+)$ &  & 5.043 & $-8.901$ & & 54.483 & $0.403$ & 0.133 & 0.307 & & 2.912 & 61.793 & $0.007$ & 0.190 & 0.002 \\
$^6$Be$(0^+)$ & $^7$Be$(3/2^-)$ & $10.677$ & $5.070$ & 2.416
 & 59.552 & $0.308$ & $0.665$ & 0.397 &  & * & * & * & * & * \\
{$^8$Be$(0^+)$} & {$^9$Be$(3/2^-)$} & $1.665$ & $-15.221$ & 2.566 & 50.094 & $0.575$ & $0.568$ & 0.496 &  & * & * & * & * & * \\
$^9$Be$(3/2^-)$ & $^{10}$Be$(0^+)$ & $6.812$ & $-12.824$ & 2.638
 & 47.421 & $1.948$ & $2.357$ & 2.147 &  & * & * & * & * & * \\
$^7$B$(3/2^-)$ & $^8$B$(2^+)$ & $12.826$ & $12.690$ & 2.413
 & 51.535 & $0.290$ & $0.725$ & 0.470 &  & 2.406
 & 51.147 & 0.103 & 0.044 & 0.127 \\
$^9$B$(3/2^-)$ & $^{10}$B$(3^+)$ & $8.437$ & $1.850$ & 2.545
 & 45.141 & $0.515$ & $0.600$ & 0.507 &  & * & * & * & * & * \\
$^8$C$(0^+)$ & $^9$C$(3/2^-)$ & $14.225$ & $12.925$ & 2.416
 & 43.600 & $0.374$ & $0.866$ & 0.699 &  & * & * & * & * & * \\
$^9$C$(3/2^-)$ & $^{10}$C$(0^+)$ & $21.284$ & $17.277$ & 2.481
 & 42.017 & $0.973$ & $1.735$ & 1.323 &  & * & * & * & * & * \\ 
\hline\hline
\end{tabular}}
\end{table*}

\begin{table*}[]
\centering
\caption{Similar to Table~\ref{SIII1}, but for proton knockout reaction.}
\resizebox{\textwidth}{!}{
\begin{tabular}{ccccccccccccccc}
\hline\hline
& & &  & \multicolumn{5}{c}{$p_{3/2}$} & & \multicolumn{5}{c}{$p_{1/2}$} \\ \cline{5-9} \cline{11-15}
$A-1$ & $A$ & $S_{p}*$ & $\Delta S$ & {$rms$} & {$\sigma_{sp}$} & $C^2S_{\rm GSM}$ & $C^2S_{\rm SM}$ & $C^2S_{\rm NCSM}$ & & {$rms$} & {$\sigma_{sp}$} & $C^2S_{\rm GSM}$ & $C^2S_{\rm SM}$ & $C^2S_{\rm NCSM}$ \\ \hline
$^6$He$(0^+)$ & $^7$Li$(3/2^-)$ & $9.974$ & $2.723$ & 2.427
 & 64.691 & $0.376$ & $0.665$ & 0.410 &  & * & * & * & * & * \\
$^7$He$(3/2^-)$ & $^8$Li$(2^+)$ & $12.416$ & $10.383$ & 2.420
 & 55.773 & $0.352$ & $0.725$ & 0.492 &  & 2.413
 & 56.274 & 0.134 & 0.044 & 0.132 \\
$^8$He$(0^+)$ & $^9$Li$(3/2^-)$ & $13.944$ & $9.882$ & 2.421
 & 49.105 & $0.453$ & $0.866$ & 0.727 &  & * & * & * & * & * \\
$^6$Li$(1^+)$ & $^7$Be$(3/2^-)$ & $5.607$ & $-5.070$ & 2.565
 & 69.453 & $0.518$ & $0.400$ & 0.469 &  & 2.583
 & 63.569 & 0.189 & 0.228 & 0.196 \\
$^6$Li$(3^+)$ &  & 7.793 & $-2.884$ &  & 69.663 & 0.426 & 0.683 & 0.430 & & * & * & * & * & * \\
$^6$Li$(0^+)$ &  & 9.170 & $-1.507$ &  & 69.738 & 0.180 & 0.333 & 0.199 & & * & * & * & * & * \\
$^6$Li$(2^+)$ &  & 9.919 & $-0.758$ &  & 69.804 & 0.039 & 0.000 & 0.057 & & 2.583 & 63.784 & 0.031 & 0.125 & 0.080 \\
$^8$Li$(2^+)$ & $^9$Be$(3/2^-)$ & $16.886$ & $15.221$ & 2.497
 & 52.118 & $0.568$ & $0.951$ & 0.687 &  & 2.485
 & 52.584 & 0.139 & 0.075 & 0.156 \\
$^8$Li$(1^+)$ &  & 17.617 & 15.952 &  & 52.114 & 0.106 & 0.055 & 0.226 & & 2.485 & 52.580 & 0.175 & 0.086 & 0.220 \\
$^9$Li$(3/2^-)$ & $^{10}$Be$(0^+)$ & $19.636$ & $12.824$ & 2.493
 & 46.163 & $1.098$ & $1.735$ & 1.356 &  & * & * & * & * & * \\
$^9$Li$(1/2^-)$ &  & 22.327 & 15.515 & * & * & * & * & * & & 2.472
 & 46.485 & 0.332 & 0.255 & 0.421     \\
{$^7$Be$(3/2^-)$} & {$^8$B$(2^+)$} & $0.136$ & $-12.690$ & 2.664 & 64.251 & $0.812$ & $1.008$ & 0.772 &  & 2.688 & 72.257 & 0.115 & 0.077 & 0.074 \\
$^7$Be$(1/2^-)$ & & 0.565 & $-12.261$ & & 56.647 & 0.278 & 0.155 & 0.230 &  & * & * & * & * & *\\
$^7$Be$(7/2^-)$ & & 4.706 & $-8.120$ & & 59.690 & 0.108 & 0.226 & 0.171 &  & * & * & * & * & * \\
$^9$Be$(3/2^-)$ & $^{10}$B$(3^+)$ & $6.587$ & $-1.850$ & 2.566
 & 48.621 & $0.530$ & $0.600$ & 0.514 &  & * & * & * & * & * \\
{$^8$B$(2^+)$} & {$^9$C$(3/2^-)$} & $1.300$ & $-12.925$ & 2.750 & 54.119 & $0.919$ & $0.807$ & 0.905 &  & 2.984 & 61.776 & 0.137 & 0.007 & 0.045 \\
$^9$B$(3/2^-)$ & $^{10}$C$(0^+)$ & $4.007$ & $-17.277$ & 2.664
 & 51.161 & $1.948$ & $2.357$ & 2.150 &  & * & * & * & * & * \\
\hline\hline
\end{tabular}}
\end{table*}

\begin{table*}[]
\centering
\caption{The results of the neutron knockout reaction in the $p$-shell have been systematically compiled. The analysis assumes that the incident nucleus collides with a $^9$Be target at an energy of 100 MeV/nucleon. Theoretical cross sections $\sigma_{th}$ are derived using the spectroscopic factors (denoted as $C^2S$) from each model, incorporating contributions from both the $p_{3/2}$ and $p_{1/2}$ partial waves. $\sigma_{sp}$ is the single-particle cross section. Here, $S_n^*$ represents the effective single-neutron separation energy, while $\Delta S$ denotes the corresponding Fermi surface asymmetry.
The rms values refer to the single-nucleon removal radii. The nuclear densities and rms values (in fm) utilized in these calculations are obtained from the Hartree-Fock-Bogoliubov method employing the SKM* interaction.}
\label{SKM1}
\resizebox{\textwidth}{!}{
\begin{tabular}{ccccccccccccccc}
\hline\hline
& & &  & \multicolumn{5}{c}{$p_{3/2}$} & & \multicolumn{5}{c}{$p_{1/2}$} \\ \cline{5-9} \cline{11-15}
$A-1$ & $A$ & $S_{n}*$ & $\Delta S$ & {$rms$} & {$\sigma_{sp}$} & $C^2S_{\rm GSM}$ & $C^2S_{\rm SM}$ & $C^2S_{\rm NCSM}$ & & {$rms$} & {$\sigma_{sp}$} & $C^2S_{\rm GSM}$ & $C^2S_{\rm SM}$ & $C^2S_{\rm NCSM}$ \\ \hline
$^7$He$(3/2^-)$ & $^8$He$(0^+)$ & $2.535$ & $-22.281$ & 2.848 & 65.412 & $3.251$ & $3.009$ & 3.317 &  & * & * & * & * & * \\
$^6$Li$(1^+)$ & $^7$Li$(3/2^-)$ & $7.251$ & $-2.723$ & 2.597 & 64.146 & $0.522$ & $0.400$ & 0.471 &  & 2.613 & 65.362 & $0.187$ & 0.228 & 0.193 \\
$^6$Li$(3^+)$ &  & 9.437 & $-0.537$ &  & 64.368 & 0.425 & 0.683 & 0.439 & & * & * & * & * & * \\
$^6$Li$(0^+)$ &  & 10.814 & $0.840$ &  & 64.290 & 0.180 & 0.333 & 0.200 & & * & * & * & * & * \\
$^6$Li$(2^+)$ &  & 11.563 & $1.589$ &  & 64.279 & $0.034$ & 0.000 & 0.058 & & 2.613 & 65.504 & $0.045$ & 0.125 & 0.079 \\
{$^7$Li$(3/2^-)$} & {$^8$Li$(2^+)$} & $2.033$ & $-10.383$ & 2.674 & 57.724 & $0.804$ & $1.008$ & 0.776 &  & 2.693 & 58.880 & $0.102$ & 0.077 & 0.066 \\ 
$^7$Li$(1/2^-)$ &  & 2.511 & $-9.905$ &  & 58.512 & 0.284 & 0.155 & 0.232 & & * & * & * & * & * \\
$^7$Li$(7/2^-)$ &  & 6.663 & $-5.753$ &  & 58.727 & 0.116 & 0.226 & 0.189 & & * & * & * & * & * \\
$^7$Li$(5/2^-)$ &  & 8.713 & $-3.703$ &  & 58.784 & 0.552 & 0.014 & 0.462 & & 2.693 & 60.172 & 0.087 & 0.001 & 0.112  \\
$^8$Li$(2^+)$ & $^9$Li$(3/2^-)$ & $4.062$ & $-9.882$ & 2.741 & 54.105 & $0.883$ & $0.807$ & 0.897 &  & 2.971 & 60.756 & $0.121$ & 0.007 & 0.041\\
$^8$Li$(1^+)$ &  & 5.043 & $-8.901$ & & 53.774 & $0.403$ & 0.133 & 0.307 & & 2.971 & 60.905 & $0.007$ & 0.190 & 0.002 \\
$^6$Be$(0^+)$ & $^7$Be$(3/2^-)$ & $10.677$ & $5.070$ & 2.490 & 60.209 & $0.308$ & $0.665$ & 0.397 &  & * & * & * & * & * \\
{$^8$Be$(0^+)$} & {$^9$Be$(3/2^-)$} & $1.665$ & $-15.221$ & 2.618 & 49.627 & $0.575$ & $0.568$ & 0.496 &  & * & * & * & * & * \\
$^9$Be$(3/2^-)$ & $^{10}$Be$(0^+)$ & $6.812$ & $-12.824$ & 2.672 & 46.947 & $1.948$ & $2.357$ & 2.147 &  & * & * & * & * & * \\
$^7$B$(3/2^-)$ & $^8$B$(2^+)$ & $12.826$ & $12.690$ & 2.484 & 52.134 & $0.290$ & $0.725$ & 0.470 &  & 2.484 & 51.750 & 0.103 & 0.044 & 0.127\\
$^9$B$(3/2^-)$ & $^{10}$B$(3^+)$ & $8.437$ & $1.850$ & 2.589 & 44.681 & $0.515$ & $0.600$ & 0.507 &  & * & * & * & * & * \\
$^8$C$(0^+)$ & $^9$C$(3/2^-)$ & $14.225$ & $12.925$ & 2.483 & 44.165 & $0.374$ & $0.866$ & 0.699 &  & * & * & * & * & * \\
$^9$C$(3/2^-)$ & $^{10}$C$(0^+)$ & $21.284$ & $17.277$ & 2.529 & 41.438 & $0.973$ & $1.735$ & 1.323 &  & * & * & * & * & *\\ 
\hline\hline
\end{tabular}}
\end{table*}

\begin{table*}[]
\centering
\caption{Similar to Table~\ref{SKM1}, but for proton knockout reaction.}
\resizebox{\textwidth}{!}{
\begin{tabular}{ccccccccccccccc}
\hline\hline
& & &  & \multicolumn{5}{c}{$p_{3/2}$} & & \multicolumn{5}{c}{$p_{1/2}$} \\ \cline{5-9} \cline{11-15}
$A-1$ & $A$ & $S_{p}*$ & $\Delta S$ & {$rms$} & {$\sigma_{sp}$} & $C^2S_{\rm GSM}$ & $C^2S_{\rm SM}$ & $C^2S_{\rm NCSM}$ & & {$rms$} & {$\sigma_{sp}$} & $C^2S_{\rm GSM}$ & $C^2S_{\rm SM}$ & $C^2S_{\rm NCSM}$ \\ \hline
$^6$He$(0^+)$ & $^7$Li$(3/2^-)$ & $9.974$ & $2.723$ & 2.501 & 64.022 & $0.376$ & $0.665$ & 0.410 &  & * & * & * & * & * \\
$^7$He$(3/2^-)$ & $^8$Li$(2^+)$ & $12.416$ & $10.383$ & 2.490 & 56.451 & $0.352$ & $0.725$ & 0.492 &  & 2.491 & 56.947 & 0.134 & 0.044 & 0.132 \\
$^8$He$(0^+)$ & $^9$Li$(3/2^-)$ & $13.944$ & $9.882$ & 2.485 & 49.729 & $0.453$ & $0.866$ & 0.727 &  & * & * & * & * & * \\
$^6$Li$(1^+)$ & $^7$Be$(3/2^-)$ & $5.607$ & $-5.070$ & 2.629 & 68.608 & $0.518$ & $0.400$ & 0.469 &  & 2.645 & 62.877 & 0.189 & 0.228 & 0.196 \\
$^6$Li$(3^+)$ &  & 7.793 & $-2.884$ &  & 68.792 & 0.426 & 0.683 & 0.430 & & * & * & * & * & * \\
$^6$Li$(0^+)$ &  & 9.170 & $-1.507$ &  & 68.857 & 0.180 & 0.333 & 0.199 & & * & * & * & * & *  \\
$^6$Li$(2^+)$ &  & 9.919 & $-0.758$ &  & 68.918 & 0.039 & 0.000 & 0.057 & & 2.645 & 63.059 & 0.031 & 0.125 & 0.080 \\
$^8$Li$(2^+)$ & $^9$Be$(3/2^-)$ & $16.886$ & $15.221$ & 2.553 & 51.560 & $0.568$ & $0.951$ & 0.687 &  & 2.552 & 52.014 & 0.139 & 0.075 & 0.156 \\
$^8$Li$(1^+)$ &  & 17.617 & $15.952$ &  & 51.555 & 0.106 & 0.055 & 0.226 & & 2.552 & 52.009 & 0.175 & 0.086 & 0.220  \\
$^9$Li$(3/2^-)$ & $^{10}$Be$(0^+)$ & $19.636$ & $12.824$ & 2.538 & 45.692 & $1.098$ & $1.735$ & 1.356 &  & * & * & * & * & * \\
$^9$Li$(1/2^-)$ &  & 22.327 & 15.515 & * & * & * & * & * & & 2.535 & 46.004 & 0.332 & 0.255  & 0.421     \\
{$^7$Be$(3/2^-)$} & {$^8$B$(2^+)$} & $0.136$ & $-12.690$ & 2.721 & 64.658 & $0.812$ & $1.008$ & 0.772 &  & 2.740 & 71.567 & 0.115 & 0.077 & 0.074 \\
$^7$Be$(1/2^-)$ & & 0.565 & $-12.261$ & & 58.740 & 0.278 & 0.155 & 0.230 &  & * & * & * & * & * \\
$^7$Be$(7/2^-)$ & & 4.706 & $-8.120$ & & 60.912 & 0.108 & 0.226 & 0.171 &  & * & * & * & * & * \\
$^9$Be$(3/2^-)$ & $^{10}$B$(3^+)$ & $6.587$ & $-1.850$ & 2.611 & 48.231 & $0.530$ & $0.600$ & 0.514 &  & * & * & * & * & * \\
{$^8$B$(2^+)$} & {$^9$C$(3/2^-)$} & $1.300$ & $-12.925$ & 2.804 & 55.222 & $0.919$ & $0.807$ & 0.905 &  & 3.067 & 63.807 & 0.137 & 0.007 & 0.045 \\
$^9$B$(3/2^-)$ & $^{10}$C$(0^+)$ & $4.007$ & $-17.277$ & 2.714 & 50.68 & $1.948$ & $2.357$ & 2.150 &  & * & * & * & * & * \\
\hline\hline
\end{tabular}}
\end{table*}